\documentclass[10pt]{bmc_article}

\usepackage{cite} 
\usepackage{url}  
\usepackage{ifthen}  
\usepackage{multicol}   
\usepackage[utf8]{inputenc} 
\usepackage{epsfig}
\newboolean{publ}

\newenvironment{bmcformat}{\fussy\setboolean{publ}{true}}{\fussy}

\begin{document}
\begin{bmcformat}

\title{Structure and evolution of protein interaction networks:
A statistical model for link dynamics and gene duplications}

\author{
        Johannes Berg\correspondingauthor$^{1}$%
        \email{Johannes Berg\correspondingauthor - berg@thp.uni-koeln.de}%
      \and
        Michael L\"assig$^1$
        \email{Michael L\"assig - lassig@thp.uni-koeln.de}
      and 
        Andreas Wagner$^2$%
        \email{Andreas Wagner - wagnera@samba.unm.edu}%
      }

\address{%
    \iid(1)Institut f\"ur Theoretische Physik,
    Universit\"at zu K\"oln, 
    Z\"ulpicherstr. 77, 50937 K\"oln, Germany\\
    \iid(2) University of New Mexico, Department of Biology, 
    167A Castetter Hall, Albuquerque, NM 817131-1091, USA
}%

\maketitle

\noindent
BMC Evolutionary Biology {\bf 4}:51 (2004)

\begin{abstract}
\paragraph*{Background}
The structure of molecular networks derives from dynamical processes
on evolutionary time scales. For protein interaction networks, global 
statistical features of their structure can now be inferred consistently 
from several large-throughput datasets. Understanding the underlying
evolutionary dynamics is crucial for discerning random parts of the 
network from biologically important properties
shaped by natural selection.

\paragraph*{Results}
We present a detailed statistical analysis of the protein interactions
in {\em Saccharomyces cerevisiae} based on several large-throughput 
datasets. Protein pairs
resulting from gene duplications are used as tracers into the evolutionary 
past of the network. ~From this analysis, we infer rate estimates for 
two key evolutionary processes shaping the network: (i) gene duplications
and (ii) gain and loss of interactions through mutations in existing proteins,
which are referred to as link dynamics. Importantly,
the link dynamics is asymmetric, i.e., the evolutionary steps are mutations
in just one of the binding parters. The link turnover is shown to be much 
faster than gene duplications. Both processes are assembled into an 
empirically grounded, quantitative model for the evolution of protein 
interaction networks. 

\paragraph*{Conclusions}
According to this model, the link dynamics is the dominant evolutionary  force
shaping the statistical structure of the network, while the slower gene
duplication dynamics mainly affects its size. Specifically, the  model predicts
(i) a broad distribution of the connectivities (i.e., the number of binding
partners of a protein) and  (ii) correlations between the connectivities of
interacting proteins, a specific consequence of the asymmetry of the link
dynamics.  Both features have been observed in the protein interaction network
of {\em S. cerevisiae}. 
\end{abstract}

\ifthenelse{\boolean{publ}}{\begin{multicols}{2}}{}


\section*{Background}
Molecular interaction networks are ubiquitous in biological systems.
Examples include transcription control \cite{gene_network}, signal
transduction, and metabolic pathways \cite{metabolic_network}.  These
networks have become a focus of recent research, because of their
important roles in metabolism, gene expression, and information
processing. Data on such networks are rapidly accumulating, massively
aided by high-throughput experiments.  Some of these networks are
sufficiently complex that their characterization requires statistical
analysis, an area of considerable recent interest
\cite{barabasi_review,dorogovtsev_review,newman_review}.  One key
issue in this area is the distinction between structures reflecting
biological function and those arising by chance. To address this issue
requires an understanding of the biological processes that shape the
network on evolutionary time scales. More precisely, one has to
identify the statistical observables containing specific information
about the evolutionary dynamics that shape a network.

In this paper we focus on protein interaction networks, whose nodes
correspond to proteins, and whose links correspond to physical
interactions between two proteins.  Several
complementary experimental techniques have been used to analyze
pairwise protein and domain interactions, as well as protein
complexes, in genome-scale assays
\cite{fromont,gavin,ho,ito,newman,rain,tong,uetz}.  Common to these
approaches is a high rate of individual false negative and false
positive interactions \cite{legrain,vonmering}. Different protein
interaction data sets thus differ in many ways, but they also reveal
similar {\it aggregate} (or global) network features, such as the
fraction of nodes with a given connectivity.  This implies that only
large-scale statistical features of protein interaction networks can
currently be reliably identified by high-throughput approaches. We
here present an empirically grounded model that explains
empirically observed statistical features of such networks.

The currently best characterized protein interaction network is that
of the baker's yeast {\it Saccharomyces cerevisiae}.  On evolutionary
time scales, this network changes through two processes, illustrated 
by figure 1. These are (i) modifications of
interactions between existing proteins and (ii) the introduction 
of new nodes and links through {\em gene duplications}. 
Duplications of a single gene result in a pair of nodes 
with initially identical binding partners. Segmental and global duplications 
of the genome lead to the simultaneous duplication of many genes. 
On the other hand, processes affecting the interactions 
between {\em existing proteins} are referred to as {\em link dynamics}. 
Link dynamics results primarily from point 
mutations leading to modifications of the interface between interacting 
proteins \cite{jones}. Both kinds of processes, link dynamics and gene duplications, 
can be inferred from a statistical 
analysis of the network data, and 
their rates can be estimated consistently with independent 
information. 

\begin{figure*}
\includegraphics[width=.65\linewidth]{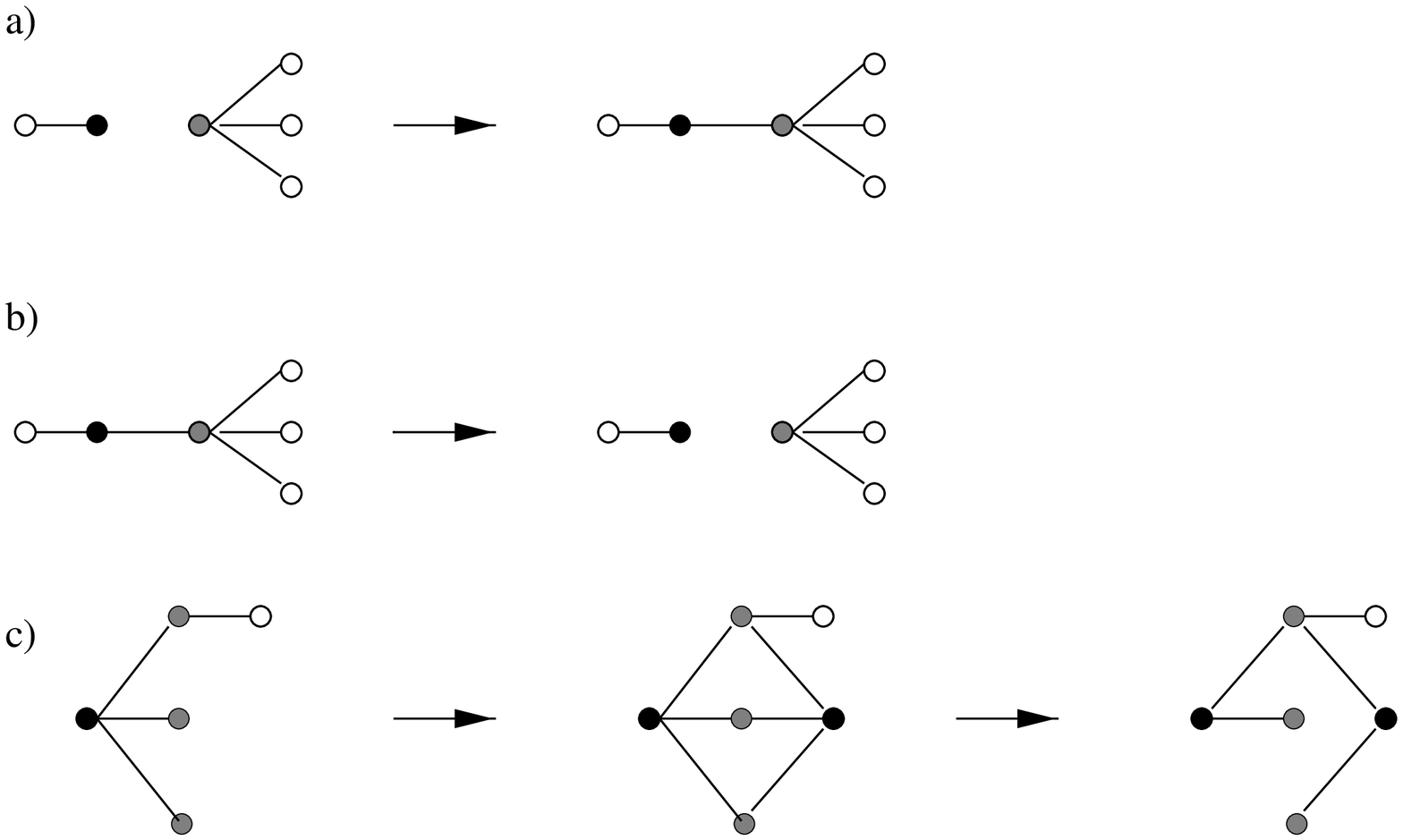}

\noindent
{\bf Figure 1: The elementary processes of protein network
evolution.} The progression of time is symbolized by arrows. 
{\bf (a) Link attachment} and
{\bf (b) link detachment}
occur through nucleotide substitutions in the gene encoding an
existing protein. These processes affect the
connectivities of the protein whose coding sequence undergoes mutation 
(shown in black) and of one of
its binding partners (shown in gray).
Empirical data shows that attachment occurs preferentially 
towards partners of high connectivity, cf. fig.~3.
{\bf (c) Gene duplication} usually produces a pair of nodes 
(shown in black) with initially identical 
binding partners (shown in gray).
Empirical data suggests duplications occur at a much lower rate 
than link dynamics and that redundant links are lost subsequently 
(often in an asymmetric fashion), which affects the connectivities
of the duplicate pair and of all its binding partners 
\protect\cite{wagner01,wagner02a,wagner02b}.     
\end{figure*}

Of course, proteome function  {\em in vivo} is influenced
by further factors, notably gene regulation, which determines the
concentrations of the proteins interacting in a living cell. The very 
definition of a bound state depends on the concentrations of the binding
partners: A pair of proteins which binds at high concentrations may no 
longer form a bound state at lower concentrations. Here we concentrate 
on protein interactions at constant concentrations as they can be inferred
from high-throughput datasets.

Previous work by others \cite{sole,vazquez,kim} shows  
how structural features of the network can in principle be explained through
mathematical models of network evolution based on gene duplications
alone. (For similar duplication-based models of regulatory and metabolic
networks, see \cite{bhan02,noort02}.)
However, the overall rate of link dynamics has  
been estimated from empirical data in \cite{wagner02b} and is 
at least an order of magnitude higher than the growth 
rate of the network due to gene duplications. It must therefore be included
in any consistent evolutionary model. 

In this paper, we present a model of network evolution that is based 
on observed rates of link and duplication dynamics. At these rates, 
the model predicts that important structural features of the network are shaped solely 
by the link dynamics. Hence, the evolutionary scenario of
our model is quite different from the duplication-based 
models~\cite{sole,vazquez,kim}. The statistical network structure predicted 
by the model is in accordance with empirical observations, see the discussion
below.  

This paper has two parts. In the first part, we estimate the rates of
link attachment and detachment from empirical data. Specifically, we
do not just estimate average rates of link dynamics for the whole
network, because this has been done previously ~\cite{wagner02b}, but
we show how the dependence of link attachment and detachment rates
depends on the connectivities of both nodes (proteins) involved.  (The
connectivity of a protein is defined as the number of its interaction
partners). We find evidence that the basic rate of link attachment is
{\em asymmetric}. That is, this rate increases with the connectivity
of only one of two the nodes involved.  This reflects an asymmetry in
the underlying biological process: a new protein-protein interaction
is typically formed through a mutation in only one of
two proteins.

In the second part of the paper, we assemble the estimated rates of
link dynamics into a model of network evolution.  Unlike for most
other cases studied so far~\cite{barabasi_review,dorogovtsev_review},
the dynamics of these networks cannot be written as a closed equation
dependent on the {\em connectivity distribution}, i.e. the fraction of
nodes with a given number of neighbors. Instead, the analysis of
networks under asymmetric link dynamics involves the {\em link
  connectivity distribution}, defined as the fraction of links
connecting a pair of nodes with given connectivities.

The model has only one free parameter, the average connectivity of
nodes in the network. Its stationary solution correctly predicts
statistical properties observed in the data.  Central properties of
this solution are {\em connectivity correlations} between neighboring
vertices, in accordance with recent observations in high-throughput
protein interaction data \cite{maslov}. These correlations are a
consequence of the asymmetric link attachment process.

\section*{Results and discussion}

\subsection*{Estimates of evolutionary rates}

Two kinds of processes contribute to the evolutionary dynamics 
of protein interaction networks. The first consists of {\em point mutations} 
in a gene affecting the interactions of the encoded protein.
As a result, the corresponding node may gain new 
links or loses some of the existing links to other nodes,
as illustrated in fig.~1(a) and~1(b), respectively. We refer to these
{\em attachment} and {\em detachment} processes, which
leave the number of nodes fixed, as {\em link dynamics}.
The second kind of process consists of {\em gene
duplications} followed by either silencing of one of the duplicated genes or by 
functional divergence of the duplicates \cite{li_book,wagner01,force}. 
In terms of the protein interaction network, a 
gene duplication corresponds to the addition of a node with links identical 
to the original node, followed by the divergence 
of some of the now redundant links between the two duplicate 
nodes; see fig.~1(c). 

Individual yeast genes have been estimated to undergo duplication at a
rate of the order of $10^{-2}$ per gene and per million years
\cite{lynch}. Some $90\%$ of {\em single} gene duplicates become silenced shortly
after the duplication, leading to an effective rate $g$ of
duplications one order of magnitude lower, i.e., $\sim 10^{-3}$ per
million years ~\cite{lynch,li,wagner01,wagner02b}.  Only a fraction of
the yeast proteome is part of the protein interaction network, and
gene duplicates involving proteins that are not part of the network do
not contribute to its growth.  Hence, $g \sim 10^{-3}$ per million
years should be considered an upper bound for the growth rate of the
protein interaction network by gene duplications.  A crude lower bound
for the link attachment rate is $a \sim 10^{-1}$ new interaction
partners per node and million years. For instance, \cite{wagner02b}
estimated the rate at which new interactions were formed as no less
than $294.5$ new interactions per million years and approximately
$1000$ proteins. (These estimates are based on the formation 
of physical interactions between products of duplicate genes, 
and the approximately known age of the duplicates \cite{wagner02b}. 
Importantly, most of these new interactions form between old duplicates, 
duplicates that are no longer under the relaxed selection pressure that 
is characteristic of young duplicates.) The above estimate gives a 
number of new interaction partners per 
protein per million years of $a=2 \times 294.5/1000=0.589$, five times
greater than the lower bound of $0.1$. To maintain an average network
connectivity at the empirically observed value $\kappa \approx 2.5$
interaction partners per protein \cite{wagner01,jeong}, the link
detachment rate $d$ has to be close to $a$, thus $d\sim a \sim
10^{-1}$ per million years.  This rate of link attachment and
detachment is much larger than the duplication rate of $g \sim
10^{-3}$ per protein and million years.  Hence, the link dynamics is
decoupled from the much slower duplication dynamics.  On intermediate
evolutionary time-scales, the network reaches a stationary state of
the link dynamics, while its number of nodes does not change
significantly. This stationary state determines the structural
statistics of the network, in particular the distribution of
connectivities. On long time-scales, however, the network may grow
through duplications.  We emphasize that all these
evolutionary rates are order-of-magnitude estimates, and that such
estimates are sufficient for our model and the conclusions we derive
from it.

One basic but important empirical observation about 
link dynamics is the fast loss of 
connectivity correlations of proteins encoded by duplicate genes. 
Fig.~2(a) shows this loss, as estimated from empirical data.
Specifically, the figure shows the average relative connectivity difference 
$|k-k'|/(k+k')$ of duplicate 
protein pairs as a function of the time since duplication, parameterized
by the fraction $K_s$ of synonymous (silent) 
nucleotide substitutions per silent site. (As an order of magnitude 
estimate, a 
value of $K_s=0.1$ corresponds to a duplication age of 
$10$ million years \cite{lynch,wagner01}.) In  
the shortest time interval after duplication, 
the connectivities are still measurably similar. Soon thereafter,
however, the relative connectivity difference 
becomes statistically indistinguishable from that of a 
randomly chosen  pair of nodes, indicated by the horizontal line 
in fig.~2(a). Hence, diversification after duplication 
is a rapid process, with a time constant of the order of several 
$10$ million years, consistent with the fast rate of link 
dynamics discussed above.  

\begin{figure*}[t!]
\includegraphics[width=.4\linewidth]{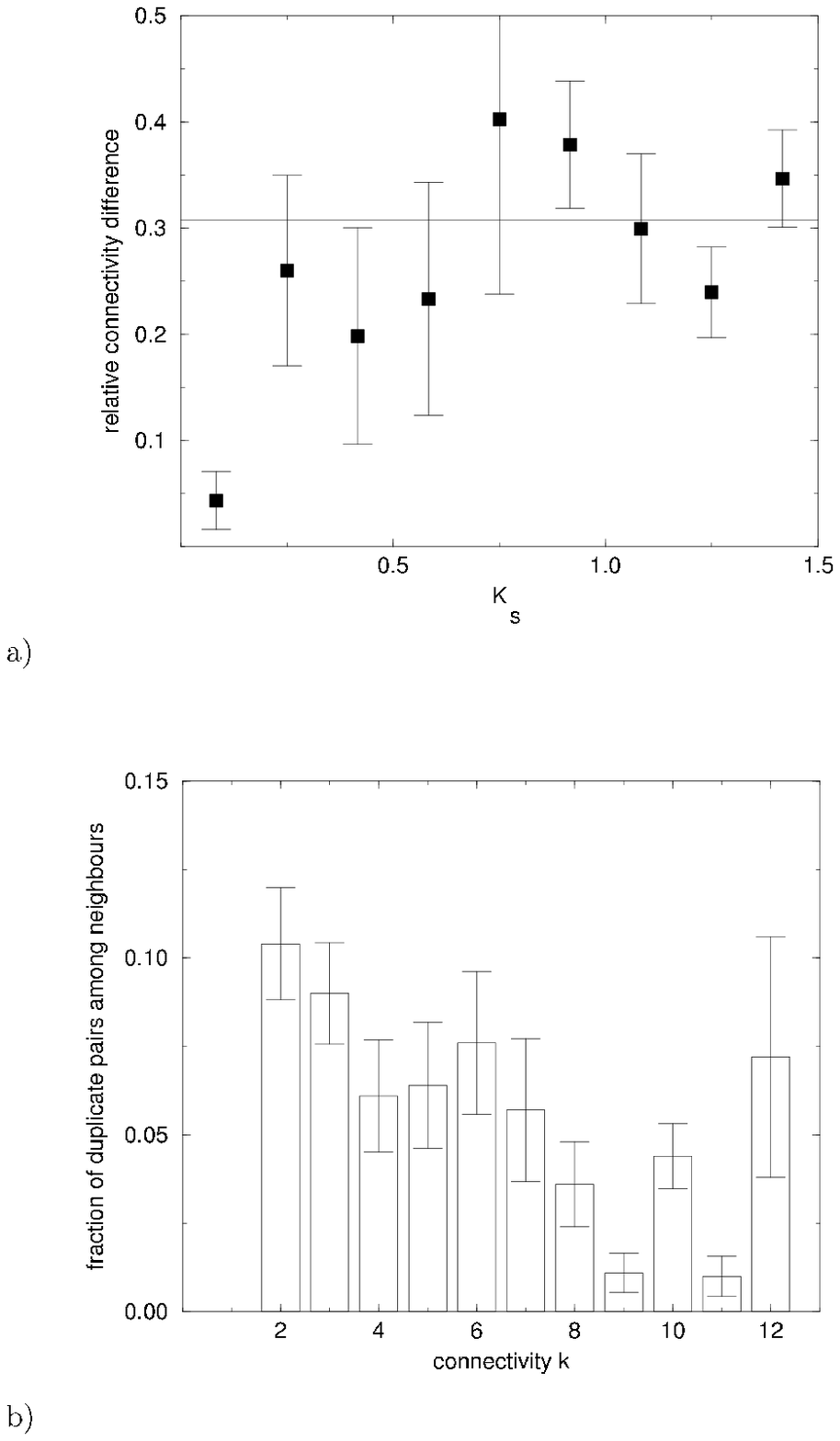}

\noindent
{\bf Figure 2(a):~ Duplicate protein pairs lose their connectivity correlations over time.}
The average relative connectivity difference $|k-k'|/(k+k')$
of duplicate pairs with connectivities $k, k'>0$  
is plotted against the time since duplication, 
parameterized by the synonymous (silent) nucleotide divergence $K_s$. 
The horizontal line indicates the value expected for two randomly chosen
nodes. The average number of duplicate pairs per bin was $16$ (from low values 
of $K_s$ to high ones the number of duplicate pairs per bin were $12,5,3,6,6,8,13,27,44$ 
respectively).
{\bf (b)~Duplications do not strongly influence network
structure.} 
The histogram shows the fraction of duplicate  
pairs among the
$k(k-1)/2$ neighbor pairs of a node of connectivity 
$k$ plotted versus $k$. 
A high number of duplicate pairs would be expected if duplications were a
significant mechanism of link gain, see text. 
The mean and the standard error of this fraction 
were determined using proteins which are products of 
duplicate genes with sequence similarity $K_a<1$.
The number of vertices used per column
ranges from $374$ for $k=2$ to $8$ for $k=12$.   
\end{figure*}

An additional 
empirical observation underscores the minor importance of gene
duplication in shaping the observed network structure.  In models of
network evolution based on gene duplication
\cite{sole,vazquez,kim}, a protein acquires new
links through duplications of its neighbors (see, for example, the
grey nodes in fig.~1(c)), at a rate proportional to its connectivity. 
This mechanism would
generate an abundance of high-connectivity nodes. 
In addition, it
would also generate a high fraction of pairs of neighbors that are
products of a gene duplication. 
This is also true for intermediate models, incorporating both 
gene duplications and link dynamics, provided the duplication 
rate is comparable to the rate of link dynamics, or exceeds it. 
However this prediction of models based on gene duplications 
is not supported by the data.  Fig. 2(b) shows the fraction of 
duplicate protein
pairs among the $k(k-1)/2$ neighbor pairs of a node of connectivity
$k$.  This fraction is small and it does not increase significantly
with $k$. The data in this figure are also consistent with the earlier
observation that the majority of duplicate pairs have few or no
interaction partners in common \cite{wagner01}.

We note that in our discussion of node dynamics we have not
separately considered the effects of ancient genome duplications~\cite{wolfe,kellis}.
The conclusion that gene duplications do not shape the statistical
features of the protein interaction network applies both to single gene
duplications and to genome duplications. Indeed, the analysis of duplicates presented
in figure~2 includes both pairs of genes resulting from single 
duplications and those stemming from genome duplications.
Furthermore, the evolutionary dynamics of individual duplicated genes
is similar for the products of single genome and whole genome duplications.
For example, individual gene duplicates are lost with
approximately the same probability in single duplications and in whole
genome duplications. For this reason we do not, at this stage, include genome
duplications separately in our model.

\subsection*{Dependency of attachment rates on connectivities}

The total rates $a$ and $d$ at which links are attached and detached
in a protein interaction network allow no inference of how these
processes shape the statistical properties of the network.  To make
such an inference, one must also know how the link dynamics depends on
the connectivities of the nodes involved. The simplest possibility is
that link attachment rates $a$ and detachment rates $d$ are functions
of a node's connectivity $k$. The rates $a_k$ and $d_k$ at which links
are attached or detached from a node of connectivity $k$ have been
estimated previously using interactions between products of duplicate
genes \cite{wagner02b}. They increase approximately linearly with $k$.

In representing attachment and detachment rates ($a$, $d$) as
functions of connectivity $k$ ($a_k$, $d_k$), one assumes implicitly
that that the mechanism of link attachment and detachment is identical 
(symmetric) for the two nodes involved in a changed link. Previous 
analyses of protein network evolution \cite{wagner02b} as well as 
models of network evolution \cite{dorosammendes} were based on 
such a symmetric process. However, the biological
mechanism underlying link dynamics is intrinsically asymmetric. When
a new link is formed, typically only one node undergoes a mutation,
whereas the other node remains unchanged. 
This asymmetry means that
the rate of link dynamics will generically depend in one 
way on the connectivity of the node undergoing mutation, and in 
another way on that of the unchanged node. As a result 
the rates $a_k$ and $d_k$ of link attachment and detachment 
are insufficient to describe the 
dynamics of the network, since these rates will be different 
depending on whether the node is undergoing a mutation or not. 
This observation motivates the following
estimate of the dependency of the link dynamics rate on node
connectivities.

We define $a_{k,k'}$ as the probability per unit time 
that a given non-interacting \emph{pair of proteins} 
with respective connectivities $k$ and $k'$ will acquire 
a link, multiplied by the number of proteins $N$. 
Analogously, we define the detachment rate $d_{k,k'}$ as the
probability per unit time that a given interacting pair of proteins
with respective connectivities $k$ and $k'$ will lose their link.
The scaling convention of both rates is chosen such 
that the average connectivity of the network remains constant 
as the number of nodes $N$ increases: the number of nodes 
pairs (where a link may be added) is proportional to $N^2$, whereas 
the total number of links (which may be deleted) is proportional to 
$N$. We refer to the special case where 
the rates factorize, i.e. $a_{k,k'} \sim a_k a_{k'}$, as symmetric 
attachment (and analogously for the detachment rates $d_{k,k'}$). 
The specific form of these rates assumes that link dynamics is a {\em local 
process}, so the probability for the formation or destruction of a link 
depends on the connectivities of only the two proteins involved in this 
process. 

We now explain how one can estimate the dependency of $a_{k,k'}$ on
its arguments, $k$ and $k'$. As described earlier \cite{wagner02b},
one can use the observed number of physical interactions among
duplicate gene products (cross-interactions) to estimate attachment
rates.  Briefly, such cross-interactions may arise in two ways.
First, a protein that forms homodimers (a self-interacting protein)
may undergo duplication, leading to two identical self-interacting
proteins which also interact with each other. If both
self-interactions are subsequently lost \emph{independently}, yet the
interaction between the nodes is retained, a cross-interaction is
formed. This scenario does probably not account for the majority of
cross-interactions, because it is inconsistent with data suggesting
that self-interactions do not get lost overly frequently after
duplication \cite{wagner02b}.  The second avenue of forming
interactions between duplicate gene products involves a
non-homodimerizing protein that undergoes duplication. Subsequently,
an interaction between the duplicate proteins may form.  If this
mechanism is dominant, as we argue, one may use the number of
cross-interactions to obtain order-of-magnitude estimates of the
attachment rate \cite{wagner02b}.  ~From the number of interactions
that each of the two involved proteins has with other proteins, one
can estimate how the attachment rate depends on $k$ and $k'$.  The
main caveat of this approach is that the connectivity of the
duplicates may have changed since the time the link between them was
formed.

The result of this procedure is shown in fig.~3. The sample size of
$38$ cross-interactions is extremely limited, but sufficient to demonstrate
an increase of the attachment rate along the diagonal $k=k'$, and no
systematic change along other directions. A different representation
of the same data in fig.~3b) also shows an increase of the attachment
rate consistent with $k+k'$.

An attachment process where one node with connectivity $k$ is chosen
with a probability $a^1_k$, and a second one is chosen with
probability $a^2_{k'}$ gives an attachment rate $a_{kk'} \sim a^1_k
a^2_{k'} + a^1_{k'} a^2_{k}$.  The attachment rate $a_{kk'} \sim k+k'$
which we observe empirically is thus explained by
an \emph{asymmetric attachment process} where one node is chosen
uniformly at random ($a^1_k=$ constant), and the other node is chosen
with a probability proportional to its connectivity ($a^2_k \sim k$).
Note that the rate $a_{k,k'}$ itself is symmetric under interchange of
the labels $k$ and $k'$, since either of the two nodes may take on the
role of being preferentially chosen. However, the rate $a_{k,k'}$ does
not factorize, exactly as required for an asymmetric attachment
process.

We now present an additional, complementary approach, based on maximum
likelihood analysis, which validates the functional form of
$a_{k,k'}$. The probability that out of $n_{k k'}$ pairs of
duplicates with given connectivities $k$ and $k'$, $m_{k k'}$ pairs
interact is $C^{n_{k k'}}_{m_{k k'}}(g_{k k'})^{m_{k k'}} (1-g_{k
  k'})^{n_{k k'}-m_{k k'}}$, where $g_{k k'}$ gives the probability
for a cross-interaction. $C^n_m = n!/\left(m! (n-m)! \right)$ are the
binomial coefficients. The probability $p$ for observing for each pair
$k \leq k'$ $m_{k k'}$ interactions in $n_{k k'}$ pairs of duplicates
is then given by $p=\prod_{k \leq k'}C^{n_{k k'}}_{m_{k k'}}(g_{k
  k'})^{m_{k k'}} (1-g_{k k'})^{n_{k k'}-m_{k k'}}$.  Symmetric and
asymmetric attachment differ in how the probability of a
cross-interaction $g_{k k'}$ depends on $k$ and $k'$. In the symmetric
case, $g_{k k'}=g_k g_{k'}$. In the asymmetric case where one node is
chosen uniformly, the other with a probability $f_k$, we have $g_{k k'}=f_k
+f_{k'}$.  Using simulated annealing~\cite{kirkpatrick} we have calculated
the (maximal) likelihoods $p$ that the connectivity correlation
pattern shown in fig.~3a resulted from either an asymmetric process,
or a symmetric process, respectively, by maximizing $p$ with respect
to $f_k$ and $g_k$.  We find that the maximal likelihood for
asymmetric attachment exceeds that for symmetric attachment
by a factor $p_{\mbox{asym}}/p_{\mbox{sym}} \sim 4$. The data thus favor
an asymmetric attachment process, consistently with the biological
interpretation given above.   In addition, in the maximum likelihood analysis
of the asymmetric model, $f_k$ shows an approximately linear increase with $k$
(see figure 3c).
Although this result is by no means conclusive, the data shows there is 
no reason to {\it a priori} consider only symmetric processes.

\begin{figure*}[t!]
\includegraphics[width=.4\linewidth]{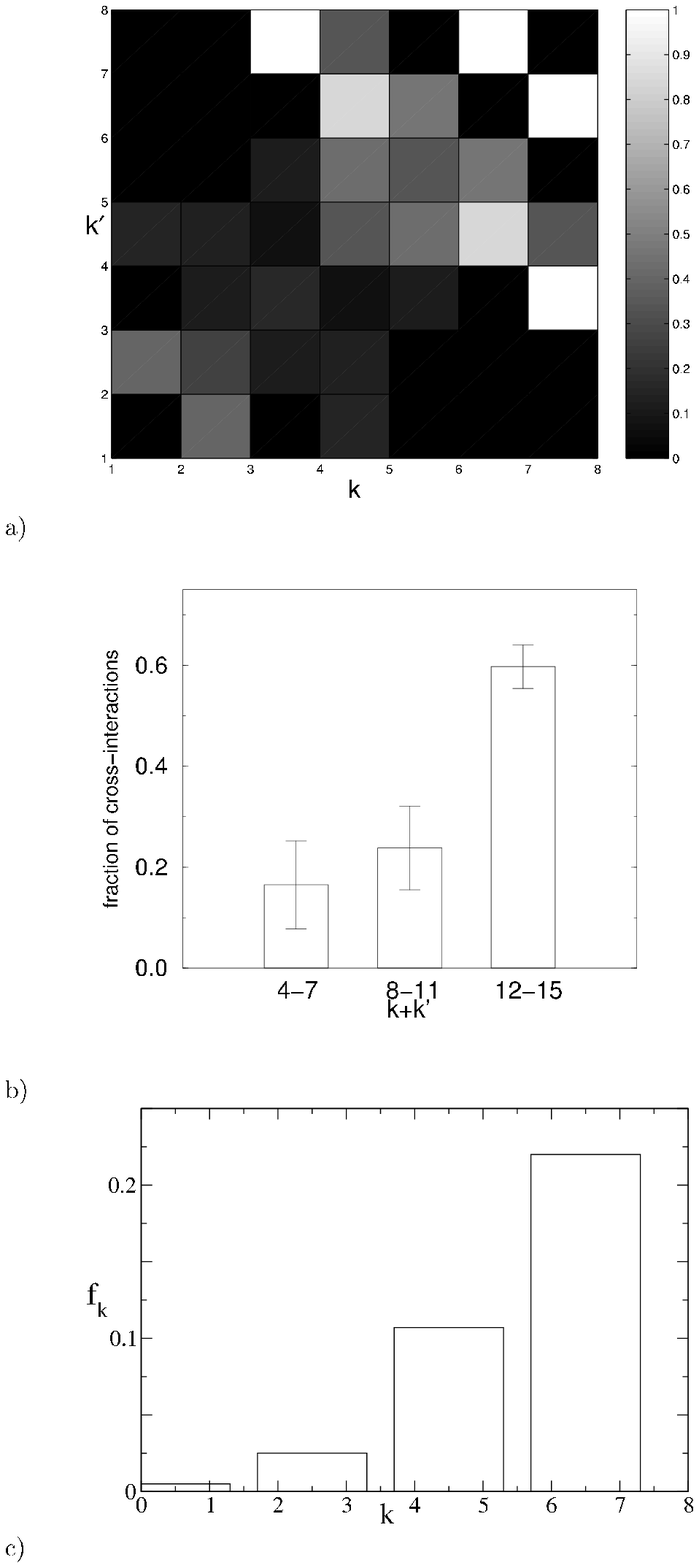}

\noindent
{\bf Figure 3: Link attachment occurs preferentially towards
proteins of high connectivity. }
{\bf (a)} The color-coded plot shows the fraction of duplicate pairs with
connectivities $(k,k')$ that have gained a mutual
interaction (cross-interaction) since duplication, as a function 
of $k$ and $k'$. 
Points where all duplicate pairs have cross-interactions are 
shown in white, points where none carry a cross-interactions are shown black. 
Points (particularly at high connectivities) where no data is available 
are also shown in black. The number of duplicate pairs with given 
connectivities ranges from $2$ to $39$. Points in the $k,k'$-plane where 
only a single pair of duplicates exists are excluded. 
{\bf (b)}
For this histogram the data from a) are binned for low, medium, and 
high $k+k'$ and the average for each bin is shown against $k+k'$. 
The number of $k,k'$ values contributing to each bin are 
$10$, $14$, and $11$, from left to right. Error bars give the 
standard error.   
{\bf (c)} Assuming the functional form $f_k+f_{k'}$ for the
probability of a cross-interaction between nodes with connectivities
$k$ and $k'$ (asymmetric attachment), the most likely values of $f_k$
may be deduced from the data (see text). The maximum-likelihood result
shows an approximately linear increase of $f_k$ with $k$. The
alternative scenario, symmetric attachment, yields a smaller maximum
likelihood. Only duplicate pairs with $K_a \leq
0.4$ were used in this analysis in order to avoid overcounting 
of cross-interactions of duplicates of even older duplicates. 
\end{figure*}

Thus far, we have only discussed the link attachment rate. For the
detachment of links, we analogously assume that links are lost due to
mutations at {\emph one} of two linked nodes, and that the rate of
this process does not depend on the properties of the other node that
is unaffected by a mutation.  The simplest mechanism reflecting these
assumptions is one where a protein loses on average $d$ links per unit
time. A protein is chosen in an equiprobable manner from all nodes for
removal of one of its links. The link to be removed is chosen at
random from all its links. (An alternative
  detachment process consists in the loss of a certain {\em fraction}
  of links and leads to very similar results.)  The resulting
detachment rate is $d_{k,k'} \sim (1/k) + (1/k')$, where the inverse
terms stem from nodes (rather than links) being chosen uniformly.

\subsection*{Dynamical model of network evolution} 

The rates of the link dynamics discussed above, together with 
a slow growth of the network due to duplications, define a simple model 
for the evolution of protein interaction networks. Unlike previous models 
of the evolution of protein interaction networks \cite{sole,vazquez,kim} 
which emphasize the 
role of gene duplications, our model is based on the asymmetric 
link dynamics deduced from empirical data in the preceding section.  
By analytical solution or 
by numerical simulation one may investigate the networks generated by our model  
and compare their statistical 
properties to those of the empirical data on protein-interaction networks. 
This will be done in the present section. 
Before analyzing this model in the limit of large networks, we discuss the 
specific values of model parameters we used, and present 
the results of numerical simulations of a finite network. 

We chose the initial network size such that after a sufficient waiting
time, when a stationary state has been reached, the size of the
simulated network matches that of the protein interaction data set we
used (see methods). Duplication of nodes is modeled simply by adding
new nodes with connectivity zero to the network at a rate of
$g=10^{-3}$ per node per million years, as motivated above.  Using
this simplistic growth mechanism is appropriate since, as shown above,
the link dynamics will quickly alter the initial connectivity of a new
node, as well as connectivity correlations with its neighbors.  We
begin with a total number of $4600$ nodes, uniformly linked at random
(giving a Poissonian connectivity distribution) such that the average
connectivity of nodes with non-zero connectivity is $\kappa=2.5$, the
average connectivity found in the data set we used.  After a waiting
time of $25$ million years there are $4696$ nodes in total, of which
$1872$ nodes have non-zero connectivity.  This is the size of the
pooled protein interaction data set we used.
The waiting time of $25$ million years is of the same order of magnitude
as  the time scale on which connectivity correlations of duplicate nodes decay 
in figure 2a) of a few $10$ million years. 

New links are added at a rate of $a=0.59$ new interactions per node
per million years, using the asymmetric preferential linking rule we
motivated above. Specifically, to form a new link we chose one node
uniformly and a second node preferentially (i.e., with a probability
proportional to its connectivity $k$) and link the two nodes.  We
removed links at a rate that keeps the average connectivity constant.
Specifically, at each time-step a link is deleted by choosing a node
uniformly for link deletion if the average network connectivity
exceeds $\kappa=2.5$. The link to be deleted is chosen equiprobably
from the links of this node. The connectivity distribution of a
network whose evolution was simulated in this manner is shown in
figure 4a) (open circles, $\circ$). This distribution is robust with
respect to changes in the ratio of duplication to link dynamics $g/a$
over at least an order of magnitude (results not shown).

\begin{figure*}[t!]
\includegraphics[width=.4\linewidth]{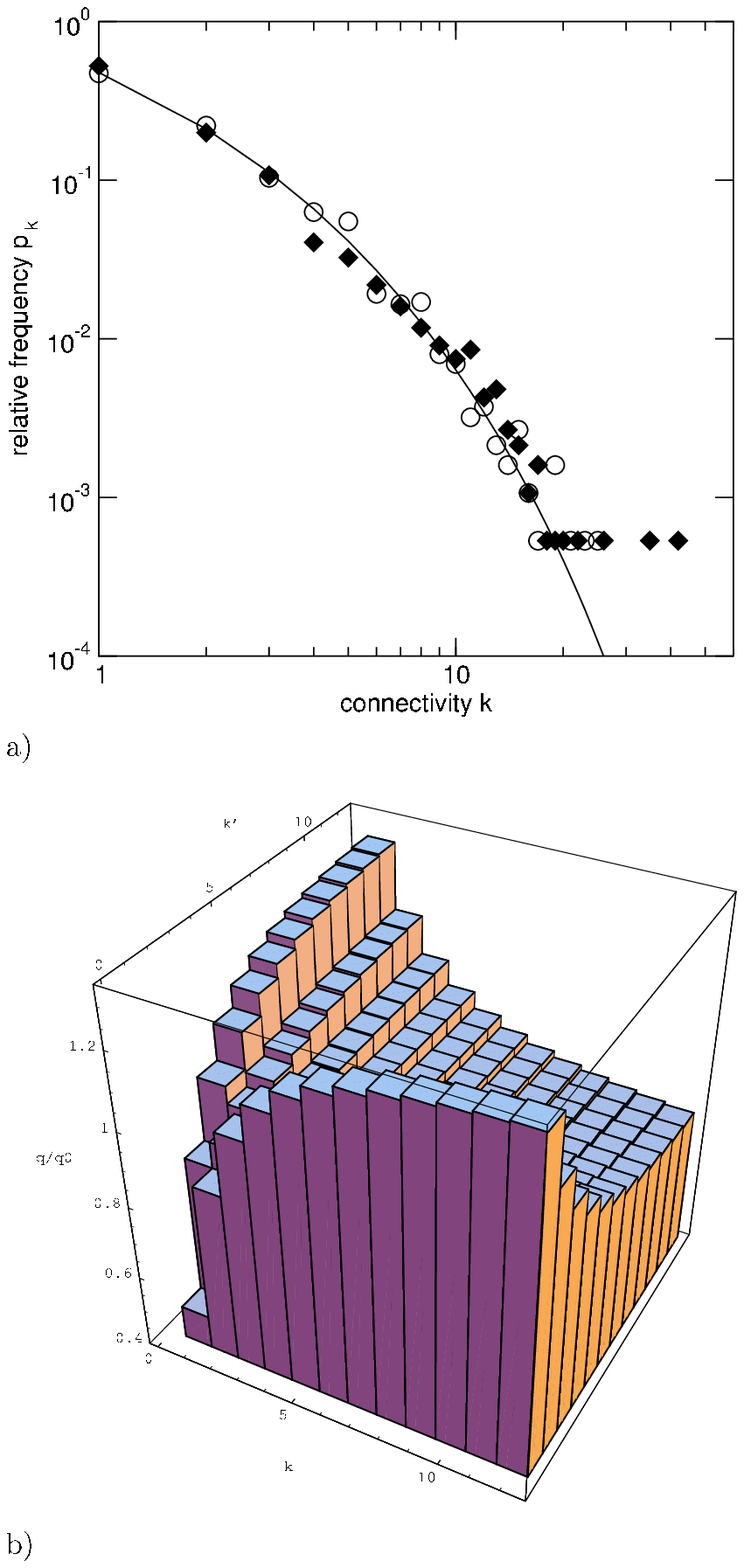}

\noindent
{\bf Figure 4(a): The asymmetric link dynamics produces a broad connectivity
distribution.}
The model prediction of the connectivity distribution of nodes with 
non-zero connectivity agrees well with yeast protein interaction data (filled diamonds). 
The solution of the rate equation (\protect\ref{qdyn}) is shown as 
a solid line, the result of a computer simulation emulating the link
dynamics encapsulated in (\protect\ref{qdyn})
for a network of finite size is shown as circles ($\circ$). 
Nodes with the highest $k$ (lower right) occur only once in the network. 
{\bf (b) High-connectivity vertices are preferentially
connected to low-connectivity vertices, as also observed empirically.}
The figure shows the relative likelihood of the
link distribution $\bar q_{k,k'}$ and the `null distribution'
$\bar q^0_{k,k'}$ of an uncorrelated random network, see text.
\end{figure*}

We now turn to the consequences of this evolutionary dynamics for the
statistical properties of the network. 
Since the link dynamics places and removes a link with a rate depending
only on the connectivities of the nodes at either end, the
evolutionary dynamics of the network can be represented in terms of the
link connectivity distribution $q_{k,k'}$. This distribution is 
defined as the fraction of network links that connect vertices  of
connectivities $k$ and $k'$,
\begin{equation}
\label{qdef}
q_{k,k'}=\frac{1}{N} \sum_{i,j} \delta_{k,k_i} c_{ij} \delta_{k',k_j} \ ,
\end{equation} 
where $c_{ij}=1$ if node $i$ is linked to $j$ and 0 otherwise.
For convenience, a factor $\kappa$ has been included in the 
normalization, i.e., $\sum_{k,k'} q_{k,k'} = \kappa$. The
link connectivity distribution $q_{k,k'}$ captures 
correlations between the connectivities of neighboring
vertices~\cite{maslov,newman_q,berg,boguna}. It is related to the
single-vertex connectivity distribution by 
\begin{equation}
p_k = \sum_{k'} q_{k,k'}/k
\end{equation}
for $k>0$ and $p_0=1-\sum_{k>0}p_k$. The 
rates  $a_{k,k'}$ and $d_{k,k'}$ are related to the total rates $a$ and $d$ 
of link detachment per unit time by the normalization 
\begin{eqnarray}
\sum_{k,k'}a_{k,k'} p_k p_{k'} &=& a \\ 
\sum_{k,k'} d_{k,k'} q_{k,k'} &=& d \nonumber \ .
\end{eqnarray}

For a network of infinite size, link and growth dynamics result in a 
deterministic differential equation for the evolution of the link 
connectivity distribution $q_{k,k'}$, 
\begin{eqnarray}
\label{qdyn}
d q_{k,k'}/dt  & = &
   a_{k-1,k'-1} p_{k-1} p_{k'-1}
   - (d_{k,k'} + g) q_{k,k'}
\nonumber \\ & &
   - (J_{k,k'} - J_{k-1,k'}) - (J_{k',k} - J_{k'-1,k}) \;.
\end{eqnarray}
The terms $J_{k,k'}$ arise from links that are not
added or removed but that change their values $(k,k')$,  
\begin{equation}
J_{k,k'} = \sum_{k''} a_{k,k''} q_{k,k'} p_{k''} - 
d_{k+1,k''+1}\frac{q_{k+1,k'}q_{k+1,k''+1}}{p_{k+1}} \frac{k}{k+1} \ .
\end{equation}
These are the links joining a mutated
protein or its binding partner with third vertices,
shown as open circles in fig.~1(a,b). The parameter $g$ accounts for 
a uniform increase of the number of nodes caused by gene duplications. 

In writing eq. (\ref{qdyn}), we have assumed that next-nearest neighbor
connectivity correlations vanish. This assumption is self-consistent since
the stationary solution has indeed only nearest-neighbor correlations. 
Truncating all correlations and writing down 
an evolution equation for the connectivity distribution 
$p_k$ turns out to be inconsistent since asymmetric link dynamics 
generates non-trivial connectivity correlations. This distinguishes the 
present model from simpler models of network growth, which can 
be self-consistently formulated at the level of the distribution $p_k$. 
  
We solved equation eq.~(\ref{qdyn}), which describes the evolution of
the connectivity correlations numerically for its steady state.  For
initial conditions we use a Poissonian connectivity distribution where
the average connectivity of connected nodes is $2.5$, and connectivity
correlations which factorize $q_{k,k'} \sim k k' p_k p_{k'}$.  We
followed the time evolution of $q_{k,k'}$ defined by eq.~(\ref{qdyn})
until a steady state was reached
using the parameters $a$ and $g$ given above and choosing $d$ such
that the average connectivity of connected nodes remains at a constant
$\kappa=2.5$.  This procedure leads to a stationary link connectivity
distribution $\bar q_{k,k'}$ and a resulting connectivity distribution
$\bar p_k$ independent of initial conditions.  
Because the evolution equation is a rate-equation that applies 
to a network of infinite size, the parameters determining the
stationary state are the ratio between growth and attachment rate, the
functional form of the attachment and detachment rates, and the
average connectivity.  The stationary state turns out to be
asymptotically independent of the duplication rate for small
duplication rates. In fact, if we solve eq.~(\ref{qdyn}) numerically
for any ratio $g/a < 10^{-1}$, the results are statistically
indistinguishable from that for $g=0$, implying great robustness
against errors in the rate estimates discussed above.

The statistical properties of our model in 
its stationary state may now be compared with the corresponding 
quantities in the protein-interaction network. 
The connectivity distribution $\bar p_k$ agrees well with the empirical 
data as shown in fig.~4(a) along with the results of numerical simulations. 
The distribution is broad but not scale
free. (~From the empirical data with connectivities
  distributed over little more than a single decade the scale-free
  property of protein networks -- meaning that connectivities are
  distributed according to a power law -- can not be confidently
  ascertained. Furthermore the empirical data shown in fig.~4
  distinctly deviates from a power-law.) This also holds for uniform
detachment, where $d_{kk'}= $ constant, and it is a crucial difference
to models with symmetric attachment, where preferential attachment
leads to scale-free networks, both at constant network
size~\cite{dorosammendes}, and in growing
networks~\cite{barabasi_review,doroinwagner}.
 
For the connectivity correlations, we find that  
vertices of high $k$ are more frequently 
linked to vertices of low $k'$ than in an uncorrelated
random network with the same connectivity distribution
$\bar{p}_k$.  Fig.~4(b) shows the relative likelihood
$\bar q_{k,k'} / \bar q^0_{k,k'}$, where 
$\bar q^0_{k,k'} = kk' \bar{p}_k \bar{p}_{k'}/\kappa$ is the link 
connectivity distribution
of the network with no connectivity correlations. 
Correlations with this property 
have recently been reported for the protein interaction  
network in yeast~\cite{maslov}, but a quantitative comparison 
with the prediction of our model will have to await 
a greater amount of reliable protein interaction data. 
We note that connectivity correlations are a specific 
property of networks shaped by asymmetric dynamics, and 
are absent in the case of symmetric dynamics, as discussed in the 
appendix.  
In other words, the empirically observed non-trivial connectivity correlations 
require asymmetric link dynamics. This is an {\em a posteriori} 
reason for considering asymmetric link dynamics. 

A further consequence of asymmetric attachment is that  
our model does not obey detailed balance (as is the case of 
symmetric link dynamics, where attachment 
and detachment rates do factorize, see \cite{dorosammendes}). 
Asymmetric attachment or detachment rules violate 
the condition, necessary for detailed balance, that 
the product of transition probabilities along a circular trajectory in 
the space of networks is independent of the direction of this tour. 
This may be demonstrated easily by considering, e.g. four nodes labeled $1-4$ 
to be connected linearly and disconnected again. Starting and ending 
with a single link between nodes $(1,2)$, say, the product of the rates 
of adding a link between $(2,3)$, then $(3,4)$ before removing the links 
between $(2,3)$ and then $(3,4)$ is $a^2_{01}d_{22}d_{11}$, that for the 
same tour in reverse is $a_{00}a_{11}d^2_{12}$, which are generally equal 
only if the rates facorize in their arguments.
 
\section*{Conclusions} 

We have presented a stochastic evolution model for protein networks,
which is based on fast link dynamics due to mutations of the coding
sequence of existing proteins and a slow growth dynamics through gene
duplications.  The crucial ingredient of the link dynamics is an
asymmetric preferential attachment rule, which is supported by
empirical data. The asymmetry has a simple biological interpretation,
namely that mutations in one gene may lead to a new interaction of its
product with that of another, unchanged, gene.  Such a mechanism,
where the two nodes involved in the generation of a new link play
different roles, is probably the norm, rather than the exception, in
biological networks. This holds particularly for regulatory networks, where 
a new interaction between two genes is formed by changes in the 
regulatory region of only {\em one} of them.

Asymmetric link dynamics leads to a
network model, where the aggregate variables necessary
to describe network structure are the connectivity correlations 
$q_{k,k'}$, which
give the fraction of links with connectivities $k$ and $k'$. In our
case, the model successfully reproduces the connectivity distribution
found in empirically available protein interaction data.  The
asymmetry of the link dynamics also leads to connectivity correlations
between interacting proteins, which have been observed empirically
\cite{maslov}.  A model with symmetric link dynamics, on the other
hand, produces no such correlations.  Higher order correlations of
this kind~\cite{berg} are of particular interest for future work as
they may be a quantitative signature of natural selection on the level
of the network as a whole. 

\section*{Methods}

\subsection*{Data processing}

The protein interaction data in this paper was pooled from three sources. 
The first of these sources is a large-scale 
high-throughput experiment using the yeast two-hybrid assay \cite{uetz} 
(data available from \cite{washington}). 
It comprises 899 pairwise interactions among 985 proteins. 
The second source is also a high-throughput two-hybrid experiment, 
from which we used a "core" set of 747 interactions between 780 proteins, 
interactions that had been confirmed through replicated experiments 
\cite{ito,kanazawa}. 
The third source is the public MIPS database \cite{mewes,mips}
of May 2001. 
~From this database, we included only pairwise interactions that 
were not produced by the two-hybrid assay, but instead by other  
techniques such as cross-linking or co-purification of two proteins. 
This resulted in 899 interactions between 680 proteins  
After pooling the three data-sets and eliminating redundant interactions, 
we were left with a network of 2463 interactions and 1893 proteins. 

While enormously valuable in their own right, analyses of protein
complexes do not identify pairwise protein interactions, and were thus
not suitable for our analysis \cite{gavin,ho}. We also excluded
interaction data derived from experiments identifying domain-specific
rather than whole-protein interactions \cite{newman,rain,tong}. For
all three data sets taken separately, the connectivity distributions
are statistically indistinguishable \cite{wagner02b}. Moreover, the
observations on link addition we use here \cite{wagner02b}, as well as
the patterns in Fig. 2 hold qualitatively for each data set
individually.

Data on yeast gene duplicates, generated as described in \cite{lynch},
was kindly provided by John Conery (University of Oregon, Department
of Computer Science).  Briefly, gapped BLAST \cite{altschul} was used
for pairwise amino acid sequence comparisons of all yeast open reading
frames as obtained from GenBank. All protein pairs with a BLAST
alignment score greater than $10^{-2}$ were retained for further
analysis. Then, the following conservative approach was taken to
retain only unambiguously aligned sequences: Using the protein
alignment generated by BLAST as a guide, a sequence pair was scanned
to the right of each alignment gap. The part of the sequence from the
end of the gap to the first "anchor" pair of matched amino acids was
discarded. The remaining sequence (apart from the anchor pair of amino
acids) was retained if a second pair of matching amino acids was found
within less than six amino acids from the first. This procedure was
then repeated to the left of each alignment gap (see \cite{lynch} for
more detailed description and justification). The retained portion of
each amino acid sequence alignment was then used jointly with DNA
sequence information to generate nucleotide sequence alignments of
genes.  For each gene pair in this data set, the fraction $K_s$ of
synonymous (silent) substitutions per silent site, as well as the
fraction $K_a$ of replacement substitutions per replacement site were
estimated using the method of Li \cite{li}. 

\subsection*{Asymmetric link dynamics and connectivity correlations}

The existence of non-trivial correlations may be attributed 
directly to the asymmetry of the link dynamics. 
Symmetric link dynamics, which is a standard mechanism in 
models of networks at constant size \cite{dorosammendes}, leads  
to networks with uncorrelated connectivities:
Generalizing the approach of~\cite{dorosammendes} to include 
connectivity-dependent detachment, one obtains for symmetric 
link dynamics with rates $a_k$ and $d_k$ an equilibrium distribution 
giving the probability of finding the network in the 
state given by adjacency matrix $c_{ij}$ of 
$P(\{c_{ij}\}) \sim \prod_{i=1}^N \prod_{k=0}^{k_i-1} a_k/d_{k+1}$.  
This results in a connectivity distribution 
$\bar{p}_k=1/(k!)\prod_{k'=0}^{k-1} a_{k'}/d_{k'+1}$ 
and trivial connectivity correlations 
$\bar q_{k,k'} \sim kk' \bar{p}_k \bar{p}_{k'}$, which factorize in the 
connectivities. This results in a constant 
$\bar q_{k,k'} / \bar q^0_{k,k'}$, where 
$\bar q^0_{k,k'} = kk' \bar{p}_k \bar{p}_{k'}/\kappa$.
A model with symmetric link dynamics can thus produce any 
empirically observed connectivity distribution, but 
no networks with statistically significant 
connectivity correlations.  

\section*{Authors' contributions}

ML and AW contributed equally to this work. 
All authors read and approved the final manuscript.

\section*{Acknowledgements}
\ifthenelse{\boolean{publ}}{\small}{}
Many thanks to S. Maslov for discussions. 
JB acknowledges financial support
through DFG grant LA 1337/1-1, AW through NIH grant
GM063882-01 and the Santa Fe Institute.
  

{\ifthenelse{\boolean{publ}}{\footnotesize}{\small}
 \bibliographystyle{bmc_article}  
  \bibliography{protdyn} }     


\ifthenelse{\boolean{publ}}{\end{multicols}}{}

\end{bmcformat}
\end{document}